# Thermal state of superconducting films on wide substrates

V. Sokolovsky, V. Meerovich[*]

Physics Depatment, Ben-Gurion University of the Negev, P. O. Box 653, Beer-Sheva, 84105, Israel

**Abstract**

The stationary thermal state and propagation of a normal zone in a long superconducting film on a wide substrate are analyzed analytically. Expressions describing voltage-current characteristics and temperature-current dependence of the film are derived for the flux creep, flux flow and normal regions. It is shown how the flux creep influences the conditions of the thermal stability. In particular, it is found that the bistability of the thermal state can appear in this regime. Under the boiling crisis, the temperature-current dependence for a film differs markedly from that for a wire and is characterized by a smooth temperature increase with the current. A "mixed" regime is analyzed where the flux flow and normal states exist simultaneously with the boundary between them parallel to the film axis. Expressions for the propagation velocity of a normal zone along a narrow film are obtained which show that this velocity in films is sufficiently higher than in wires.

1. **Introduction**

State-of-the-art technology allows one to produce large-size high–temperature superconducting (HTS) and $MgB_2$ superconducting films and coated superconductors with a high critical current density [1-3]. One of the most important problems in applying superconductors is their thermal stability. In all the cases, the thermal state of a film or a coated superconductor is determined by the competition between the losses and the heat removal where the substantial role plays the substrate. To investigate the thermal stability, two types of the problems are usually considered. The first type is related to a thermal disturbance due to energy release in an infinite line source along a superconductor [4]. The second type of problems is related to heat conduction and normal zone propagation along a superconductor [5-7]. To solve these problems, it is frequently assumed that temperature of the film and substrate changes only along the film and does not vary in the direction across the film. This approximation is correct when the width of a film and a substrate is much less than the characteristic thermal length which can be estimated as [5]

$$l_t = [\lambda(T_0)\Delta /H(T_0)]^{1/2},$$

where $\lambda$ is the thermal conductivity of the structure; $H$ is the heat transfer coefficient to coolant; $T_0$ is the temperature of the coolant; $\Delta$ is the thickness of the substrate. For the sapphire substrate this length is about 1 cm at 77 K. This value is order of a substrate width for electronic applications and much less than the substrate width for power engineering applications and, what is more, the superconducting film width can achieve of this value [1].

[*] Corresponding author: Tel.: +972 8 647 2458; fax: +972 8 647 2903.
E-mail address: victorm@bgu.ac.il (V. Meerovich).

Numerical simulations were mainly devoted to analysis of the formation and propagation of normal zones [8-10]. These investigations are based on the consideration of two- and three-dimensional models with various *E-J* relationship of a superconductor. However, it is difficult from their results to determine voltage-current characteristics (VCC) of a superconductor and the practically important parameters such as the minimal currents of propagation and existence of a normal zone, the quenching current, and etc.

In this paper we present analytical expressions describing the thermal state of a film deposited on a wide substrate. The obtained expressions allow one to analyze the influence of the substrate width and determine the ranges of the thermal stability of superconducting films.

## 2. Mathematical Model

The thermal state of a superconducting film on a thin substrate can be described by the equation

$$C\Delta \frac{\partial T}{\partial t} = \left[ \frac{\partial}{\partial x}\left(\lambda \Delta \frac{\partial T}{\partial x}\right) + \frac{\partial}{\partial y}\left(\lambda \Delta \frac{\partial T}{\partial y}\right) \right] + W(J,T) - H(T)(T-T_0) \quad (1)$$

where $T$ is the temperature; $C$ is the specific heat capacity of the substrate; $W(J,T)$ is the losses in the film per unit of the film surface, $J$ is the local current density. Here we assume that the heat capacity and thermal conductivity of the composite are determined by the thermal parameters of the substrate.

The boundary conditions for Eq. (1) are determined by the continuity of the temperature and heat flux at the film boundaries and by neglecting a heat flux from face planes of the substrate.

The losses in a superconductor depend on a local current density $J$ and temperature $T$. At a temperature below the critical value $T_c$ the losses can be caused by flux creep (FC) or flux flow (FF). In low temperature superconductors (LTS) the FC regime is observed only in very narrow range of currents near the critical value $I_c$ and causes an electric field of about 1 μV/cm. Losses in this regime are usually neglected at the analysis of the thermal stability of a superconductor. As distinct from LTS, the FC regime in HTS is observed at currents far less than the critical value and the electric field can achieve 1 mV/cm [11]. Such a regime is accompanied by a pronounced heating of a superconductor and can lead to quenching [6, 11]. The *E-J* relation in the FC regime is frequently fitted by a power low $E = E_0 (J/J_0)^n$. One of the parameters $E_0$ and $J_0$ of the fitting is chosen arbitrarily. Therefore we assume $J_0$ to be equal to critical current density $J_c$ determined as the dividing value between the FC and FF regimes. The critical current and its density are assumed to be linearly dependent on the temperature and proportional to $1-(T-T_0)/(T_c-T_0)$.

The voltage drop across a superconductor in the FF regime and in the normal state can be approximated by linear functions of a current and a temperature. Thus, the voltage drop per unit length of a superconductor in different ranges of the current and temperature can be presented as

$$U = \begin{cases} U_1 [J/J_c(T)]^n & \text{if } J < J_c(T) \\ U_1 + \rho(T_c)[J - J_c(T)] & \text{if } J \geq J_c(T) \text{ and } T < T_c \\ \rho(T_c)[1+\eta(T-T_c)]J & \text{if } T \geq T_c \end{cases} \quad (2)$$

where $\rho$ is the film resistivity in the normal state; $\eta$ is the temperature coefficient of the resistivity. Note that this approximation gives jumps in curves of voltage and of temperature at $T$

= $T_c$. However, in many practical cases $U_1 \ll \rho(T_c)J_c$ and this jump can be neglected. The jump can be deleted by introducing a dependence of $U_1$ on temperature so as $U_1(T_c) = 0$. Below we consider two cases:

$$U_1 = U_0 = \text{const} \quad \text{and} \quad U_1 = U_0\left(1 - \frac{T-T_0}{T_c - T_0}\right). \tag{3}$$

Eq. (2) well fits also VCCs of superconducting films covered by a normal well-conducting layer. In this case the resistivity in Eq. (2) is an equivalent resistivity taking the ratio of the thicknesses of the superconducting and normal conducting layers.

The integration of the current density over the film cross-section gives the total current $I$:

$$I = \Delta_f \int_{-L}^{L} J d\xi \tag{4}$$

where $L$ is a half of the film width; $\Delta_f$ is the thickness of the film.

Assuming that the parameters $C$ and $\lambda$ are independent of temperature, Eqs. (1)-(4) can be rewritten in the following dimensionless form:

$$\frac{\partial \tau}{\partial t^*} = \frac{\partial^2 \tau}{\partial x^{*2}} + \frac{\partial^2 \tau}{\partial y^{*2}} + \alpha u j - h\tau \tag{5}$$

$$u = \begin{cases} u_1 \left(\dfrac{j}{1-\tau}\right)^n & \text{if } j < 1-\tau \\ u_1 + j - 1 + \tau & \text{if } j \geq 1-\tau \text{ and } \tau < 1 \\ [1 + \gamma(\tau - 1)]j & \text{if } \tau \geq 1 \end{cases} \tag{6}$$

$$u_1 = u_0 \quad \text{or} \quad u_1 = u_0(1-\tau) \tag{7}$$

$$2li = \int_{-l}^{l} j d\xi \tag{8}$$

where $j = J/J_c(T_0)$; $i = I/I_c(T_0)$; $I_c(T_0) = 2LJ_c(T_0)\Delta_f$ is the critical current of the film at $T = T_0$; $\tau = (T-T_0)/(T_c-T_0)$; $t^* = tH(T_0)/C(T_0)\Delta$; $\gamma = \eta(T_c-T_0)$; $u_0 = U_0/\rho(T_0)J_c(T_0)$; $x^* = x[H(T_0)/\lambda(T_0)\Delta]^{1/2}$; $l = L[H(T_0)/\lambda(T_0)\Delta]^{1/2}$; $h = H(T)/H(T_0)$; $y^* = y[H(T_0)/\lambda(T_0)]^{1/2}$; $\alpha = \rho(T_0)J_c(T_0)^2\Delta_f/[H(T_0)(T_c-T_0)]$ is the analog of so called Stekly's parameter. Henceforth, the symbol "*" is omitted in the notations of the dimensionless values.

## 3. Stationary thermal state

Let us analyze the thermal state of an infinite long straight superconducting film directed along the x-axis (Fig. 1). The film is symmetrically deposited on a substrate of the width $2l_s$. Eq. (5) can be represented for a homogeneous film in the stationary state in the following form:

$$\frac{d^2\tau_f}{dy^2} + \alpha u j - h\tau_f = 0 \quad \text{for } -l < y < l; \tag{9}$$

$$\frac{d^2\tau_1}{dy^2} - h\tau_1 = 0 \qquad \text{for } l \leq |y| \leq l_s; \tag{10}$$

where $\tau_f$ and $\tau_1$ are the temperatures in the area occupied by the film and outside the film, respectively. (The temperatures of a film and a substrate are the same in the area occupied by the film).

Due to the task symmetry we will consider only a half of the composite ($y \geq 0$). The boundary conditions for Eqs. (9) and (10) are

$\tau_f = \tau_1$ and $d\tau_f/dy = d\tau_1/dy$ for $y = l$; (11)
$d\tau_1/dy = 0$ for $y = l_s$; (12)
$d\tau_f/dy = 0$ for $y = 0$. (13)

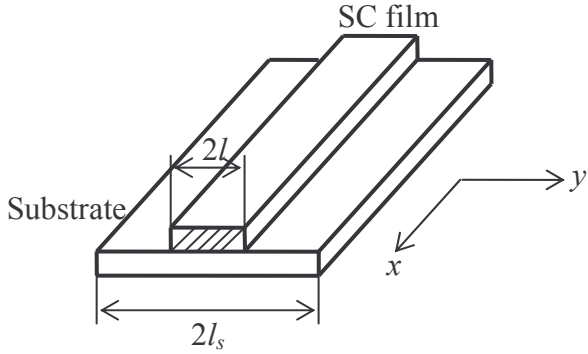

Fig. 1. Sketch of a SC film on a wide substrate.

*3.1. Narrow film*

At $l \ll 1$ the temperature across the film can be taken constant so that a current density $j$ is independent of $y$. Integration of Eqs. (8) and (9) gives:

$j = i$
and

$$hl\tau_f = \left.\frac{d\tau_1}{dy}\right|_{y=l} + l\alpha ui. \tag{14}$$

In the case when the heat transfer coefficient is independent of temperature, $h(\tau)=1$, the solution of Eq. (10) with boundary conditions (11) and (12) gives

$$\tau_1 = \tau_f \cosh(l_s - y)/\cosh(l_s - l) \tag{15}$$

From Eqs. (14) and (15) we obtain the following expression for the temperature of a superconductor:

$$\tau_f = \alpha_{ef} ui, \tag{16}$$

where $\alpha_{ef} = \dfrac{\alpha l}{l + \tanh(l_s - l)}$. (17)

Expression (16) is similar to the equation usually used for investigations of the thermal state of a superconductor (see, for example, [12,13] and references noted in them). Here the effective parameter $\alpha_{ef}$ is used instead of the Stekly parameter $\alpha$. In the denominator of (17) the first term corresponds to a heat flux from the film surface to a coolant, the second one – to a heat flux from the film through the substrate. It is seen that an increase of value $l_s$-$l$ above 2 does not lead to a marked change of the parameter $\alpha_{ef}$. Therefore further widening the substrate does not increase the thermal stability of the film. At $l \to 0$ the product $\alpha l$ remains a finite value characterizing losses in the film, and the cooling of the film is realized only through the substrate.

Using Eqs. (6) and (16) one can build VCCs of the film and determine the characteristic values of the current which important for analysis of the thermal stability. Tabl. 1 and 2 summarize basic relationships between current, temperature and voltage drop in FC and FF regimes, respectively. Tabl. 1 includes also the expressions for the maximum temperature $\tau_{fm}$ in the FC regime and corresponding values of the voltage $u_m$ and current $i_1$. In the normal state ($\tau_f \geq 1$) the film temperature and the voltage drop are determined by the following expressions:

$$u = \frac{i(1-\gamma)}{1-\gamma \alpha_{ef} i^2}$$

$$\tau_f = \frac{\alpha_{ef} i^2 (1-\gamma)}{1-\gamma \alpha_{ef} i^2}.$$

One can see that there is the maximum current of the normal state $i_m = 1/(\alpha_{ef}\gamma)^{1/2}$ above which the temperature increases unlimitedly and the film burns. The minimal current of the existence of the normal state is determined as for a wire: $i_{min} = 1/\sqrt{\alpha_{ef}}$.

Figs. 2 and 3 present VCCs and temperature-current dependences for two considered cases of $u_1$ according to (7) with $u_0 = 0.1$ and $\gamma = 0.1$. It is seen that both models give qualitatively close results.

At the analysis of the thermal stability of superconductors, it is frequently assumed that $u_0 \ll 1$ and the voltage drop in the FC regime can be neglected. This approximation well works for LTS, however can lead to pronounced errors for HTS. The VCCs of LTS have been discussed in details in many publications (for example see [12, 13]). The main difference between VCCs of HTS and of LTS wires lies in the existence of the region of FC regime. Fig. 4 presents the details of VCCs in the FC regime (zoom of Fig. 3a). There are two types of curves: (1) $u(i)$ is a single-valued function ($\alpha_{ef} = 0.5$ and $\alpha_{ef} = 1.235$ in Fig. 4); (2) $u(i)$ is a double-valued function in the range $i_1 < i < i_2$ ($\alpha_{ef} = 3$ and $\alpha_{ef} = 5$). The current $i_2$ has a meaning of the maximum current of the existence of the FC regime. The expressions for $i_2$ are also presented in Tabl. 2.

The values of $\alpha_{ef} u_0$ and $n$ determine what type of the VCC will be realized. The boundary value of $\alpha_{ef} u_0$ between two types of curves can be determined from the condition: $i_1 = i_2$ that gives $(\alpha_{ef} u_0)_{cr} = 1/n$ for $u = u_0 =$ const, and $(\alpha_{ef} u_0)_{cr} = n/(n-1)^2$ for $u = u_0(1-\tau_f)$. If $\alpha_{ef} u_0$ is less than the boundary value, VCC is a single-valued function; above -VCC has an ambiguous part. In the last case the curve has a dropping part and, in the range $i_1 < i < i_2$, there is a region of bistability. The second stable state in this region is the normal state (Figs. 2 and 3).

The obtained VCCs can explain the observed difference between the critical current $i_{cmes}$ measured by the pulse method and the quench current $i_q$ measured at applying DC: $i_q < i_{cmes}$ [14]. At a short current pulse, heating is negligible and the critical current $i_{cmes}$ is determined from the VCC obtained at the coolant temperature (dashed line in Fig. 4). At the DC conditions, heating in the FC regime can markedly increase the temperature of the film. Further scenario depends on $\alpha_{ef}$. For example, for $\alpha_{ef} = 5$ only the normal state exists in the stationary regime within the current range $i_1 < i < i_{cmes}$ (Fig. 3). Therefore any current in this range causes the quenching. Fig. 5 shows the nucleation and development of the normal zone in a HTS film deposited on sapphire substrate [14]. The critical current obtained by a pulse method was about 32 A. The quench starts at about 20% less than this value (26 A).

Table 1

Basic relationships between current, temperature and voltage in the FC regime for cases $u_1 = u_0$ and $u_1 = u_0 (1-\tau_f)$

| $u_1 = u_0$ | $u_1 = u_0 (1-\tau_f)$ |
|---|---|
| $i = \dfrac{(u/u_0)^{1/n}}{1 + \alpha_{ef} u_0 (u/u_0)^{1+1/n}}$ | $i = \left[ \dfrac{\tau_f (1-\tau_f)^{n-1}}{\alpha_{ef} u_0} \right]^{1/(n+1)}$ |
| $\tau_f = \dfrac{\alpha_{ef} u_0 (u/u_0)^{(n+1)/n}}{1 + \alpha_{ef} u_0 (u/u_0)^{(n+1)/n}}$ | $\dfrac{u}{u_0} = \left[ \left( \dfrac{\tau_f}{\alpha_{ef} u_0} \right)^n \dfrac{1}{(1-\tau_f)^{n-1}} \right]^{1/(n+1)}$ |
| $\tau_{fm} = \dfrac{\alpha_{ef} u_0}{1 + \alpha_{ef} u_0}$ | $\tau_{fm} = \dfrac{1 + 2\alpha_{ef} u_0 - \sqrt{1 + 4\alpha_{ef} u_0}}{2\alpha_{ef} u_0}$ |
| $u_m = u_0$ | $\dfrac{u_m}{u_0} = \dfrac{\sqrt{1 + 4\alpha_{ef} u_0} - 1}{2\alpha_{ef} u_0}$ |
| $i_1 = 1/(1+\alpha_{ef} u_0)$ | $i_1 = \dfrac{\sqrt{1 + 4\alpha_{ef} u_0} - 1}{2\alpha_{ef} u_0}$ |
| $i_2 = \dfrac{n}{(\alpha_{ef} u_0 n)^{1+1/n} (n+1)}$ | $i_2 = \left[ \dfrac{(n-1)^{n-1}}{n^n \alpha_{ef} u_0} \right]^{1/n+1}$ |

Table 2

Basic relationships between current, temperature and voltage in the FF regime for cases $u_1 = u_0$ and $u_1 = u_0 (1-\tau_f)$.

| $u_1 = u_0$ | $u_1 = u_0 (1-\tau_f)$ |
|---|---|
| $u = \dfrac{u_0 + i - 1}{1 - \alpha_{ef} i}$ | $u = \dfrac{u_0 + i - 1}{1 - \alpha_{ef} i (1 - u_0)}$ |
| $\tau_f = \dfrac{\alpha_{ef} i (u_0 + i - 1)}{1 - \alpha_{ef} i}$ | $\tau_f = \dfrac{\alpha_{ef} i (u_0 + i - 1)}{1 - \alpha_{ef} i (1 - u_0)}$ |

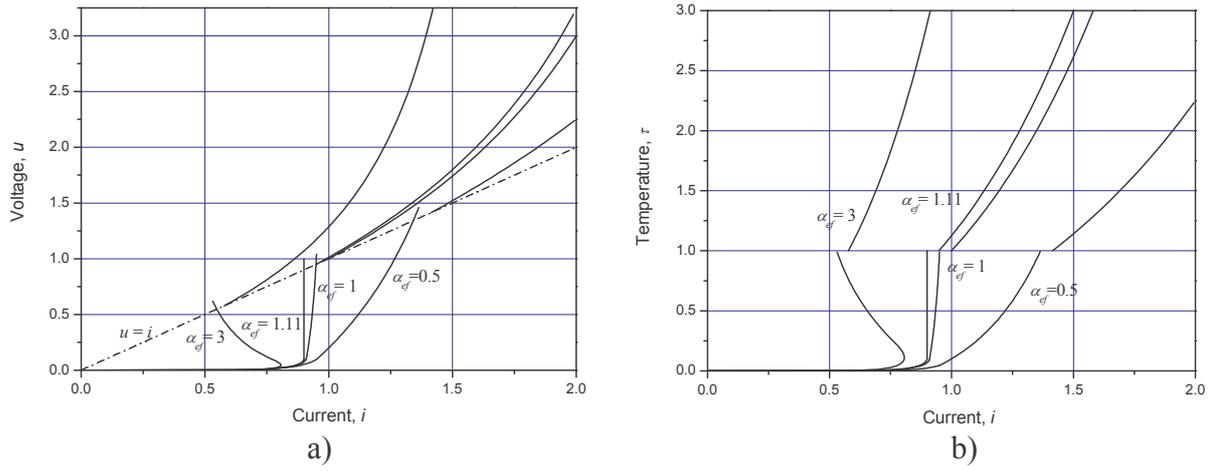

Fig. 2. VCC (a) and temperature-current dependence (b) for $u_1$=constant.

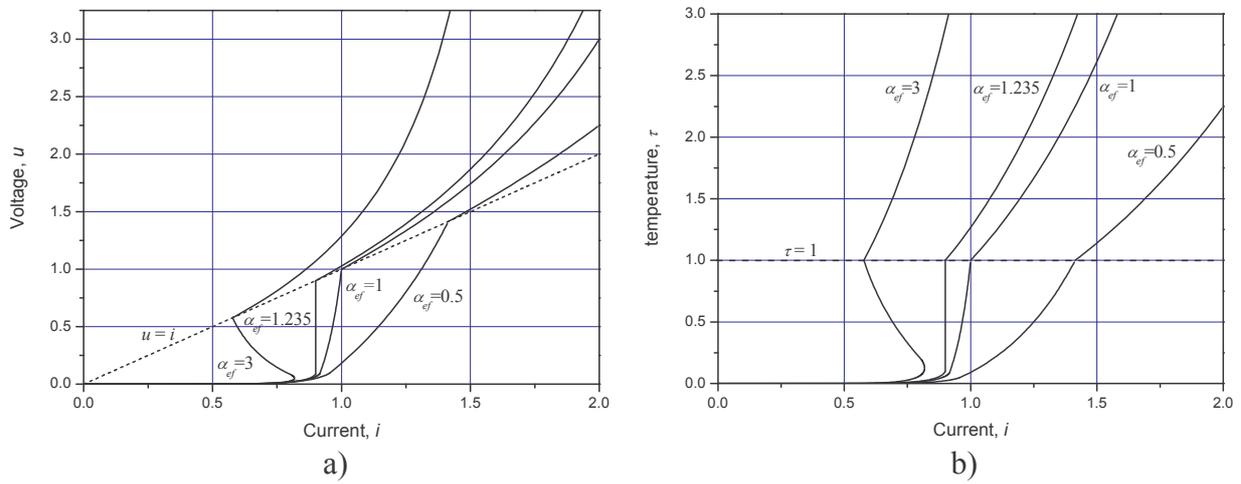

Fig. 3. VCC (a) and temperature-current dependence (b) for $u_1=u_0(1-\tau)$.

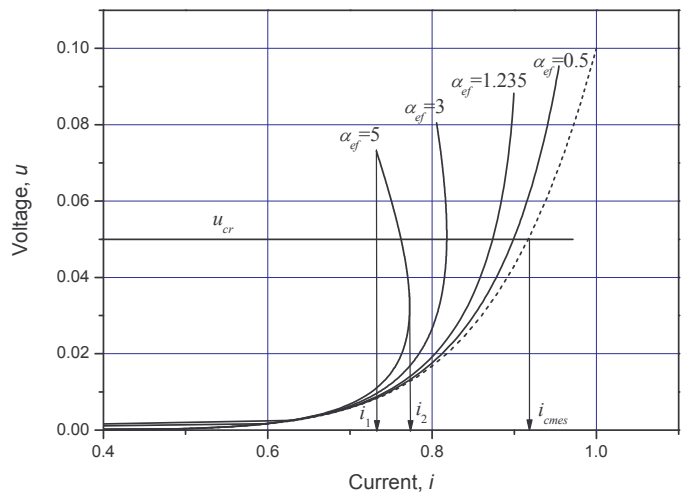

Fig. 4. VCCs in FC regime for $u_1 = u_0(1-\tau_f)$ and $u_0 = 0.1$. Dashed curve presents VCC without account of heating.

The FF regime starts at the current $i_2$. Since the thermal state is unstable on the negative slop of a VCC, there are curves with stable ($\alpha_{ef} < \alpha_c$) and unstable ($\alpha_{ef} > \alpha_c$) states in the FF regime (Figs. 2 and 3). For LTS, the boundary separated the stable and unstable regions is a vertical line at current $i = 1$ with $\alpha_c = 1$ [12]. For HTS, this vertical line is obtained at $\alpha_c = 1/(1-u_0)$ for $u_1 =$ const and at $\alpha_c = 1/(1-u_0)^2$ for $u_1 = u_0(1-\tau_f)$ and at current $i = 1-u_0$ for both cases (for example, a vertical line at $\alpha_{ef} = 1.235$, Fig. 3a).

Note that the described features introduced by the FC are also related to SC wires.

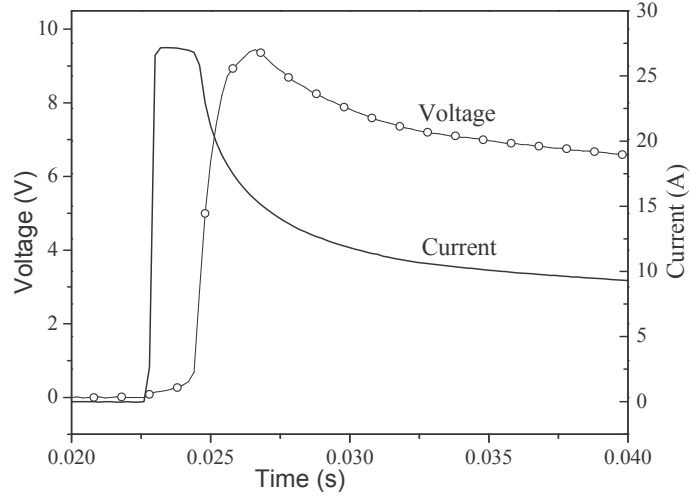

Fig. 5. Nucleation and development of normal zone in a thin film under DC conditions. Shown: voltage drop across the film section and current in the film.

*3.2. Influence of boiling crisis*

The critical temperature of an HTS can be either above or below the boiling crisis temperature of the coolant $\tau_{cr}$. Therefore, boiling crisis can be observed in all the states: FC, FF and normal. To investigate the influence of the boiling crisis on the stability of a superconductor, let us consider the case $\tau_{cr} > 1$ allowing one to obtain an analytical solution. The boiling crisis is frequently simulated by a jump of the heat transfer coefficient from 1 down to $h_1$ at $\tau = \tau_{cr}$. Let $\gamma = 0$, $l_s \gg 1$ and $l \ll 1$. Then the temperature of the substrate is determined from Eq. (10) with the boundary conditions

$$\alpha l i^2 = -\left.\frac{d\tau_1}{dy}\right|_{y=0}$$
$$\tau_1(l_c) = \tau_{cr} \qquad\qquad (18)$$
$$\left.\frac{d\tau_1}{dy}\right|_{y=l_c-0} = \left.\frac{d\tau_1}{dy}\right|_{y=l_c+0}.$$

The solution of Eq. (10) is

$$\tau_1 = \begin{cases} A\exp\left(-\sqrt{h_1}\,y\right) + B\exp\left(\sqrt{h_1}\,y\right) & 0 \leq y \leq l_c \\ \tau_{cr}\exp(l_c - y) & y > l_c \end{cases}, \quad (19)$$

The coordinate $y = l_c$ separating the areas of the film boiling at $y < l_c$ and the nuclei boiling at $y > l_c$ is determined from the boundary conditions (18). The equation for $l_c$ is

$$\sqrt{h_1}\,\tau_{cr}\sinh\left(\sqrt{h_1}\,l_c\right) - \alpha l i^2 = \cosh\left(\sqrt{h_1}\,l_c\right)$$

The equation has two solutions but only one has a physical meaning:

$$l_c = \frac{1}{\sqrt{h_1}} \ln\left[\frac{\alpha l i^2 / \tau_{cr} + \sqrt{\left(\alpha l i^2 / \tau_{cr}\right)^2 - 1 + h_1}}{1 + \sqrt{h_1}}\right] \quad (20)$$

The film boiling appears at $y = 0$ when the film temperature achieves $\tau_{cr}$. The width of the film boiling zone slowly increases with current according to the logarithmic law. A clear boundary between the zones of the film and nuclei boiling was observed in some experiments (see, for example the photos presented in [1]).

In the approximation $l \ll 1$ the film temperature $\tau_f$ is equaled to $\tau_1$ at $y = 0$:

$$\tau_f = \frac{\tau_{cr}}{\sqrt{h_1}} \sqrt{\left(\frac{\alpha l i^2}{\tau_{cr}}\right)^2 - 1 + h_1} \quad (21)$$

Fig. 6 presents the dependence of the temperature in the normal state ($\tau > 1$) on the current for a film (solid line) and a wire (dashed line). The Stekly parameter for the wire $\alpha_s$ was chosen equal to the corresponding parameter $\alpha l$ for the film. Therefore both curves coincide before the boiling crisis point. The crisis takes place at current $i = \sqrt{\tau_{cr}/\alpha l}$. The further behavior differs for a film and a wire. While the temperature of a wire jumps from $\tau_{cr}$ up to $\tau_{cr}/h_1$, the temperature of a film on a wide substrate increases gradually.

Note that in models involving the boiling crisis it is impossible to introduce an effective parameter $\alpha_{ef}$. Only for $(\alpha l i^2/\tau_{cr})^2 \gg 1$, $\tau_f \approx \alpha l i^2/\sqrt{h_1}$ and we can use the effective parameter $\alpha_{ef} = \alpha l/\sqrt{h_1}$, which increases $1/\sqrt{h_1}$ times at the boiling crisis. Remember that for wires and cables the Stekly parameter is increased $1/h_1$ times [12].

For both a film and a wire, the decrease of the current leads to gradual reduction of the temperature. As a result the temperature-current dependence for a wire has a hysteretic character and the film boiling is observed till a current $i_{mbw} = (h_1 \tau_{cr}/\alpha_s)^{1/2}$. In the case shown in Fig. 6, this minimum current of the film boiling is at the same time the minimum normal zone existence current. Note that at the point $i_{mbw}$ the temperature is $\tau_{cr}$ and above the critical one ($\tau = 1$). As distinct from it, in the case of a film, the return occurs along the forward curve and the film boiling passes smoothly to the nuclei boiling. The minimum normal zone existence current is determined as $i_{min} = 1/\sqrt{\alpha l}$ independent of $h_1$ and $\tau_{cr}$. At the parameters of the case of Fig. 6, the ratio $i_{min}/i_{mbw}$ is about 2.

Thus, the comparison shows that the thermal stability of a film deposited on a wide substrate is higher than the stability of wires.

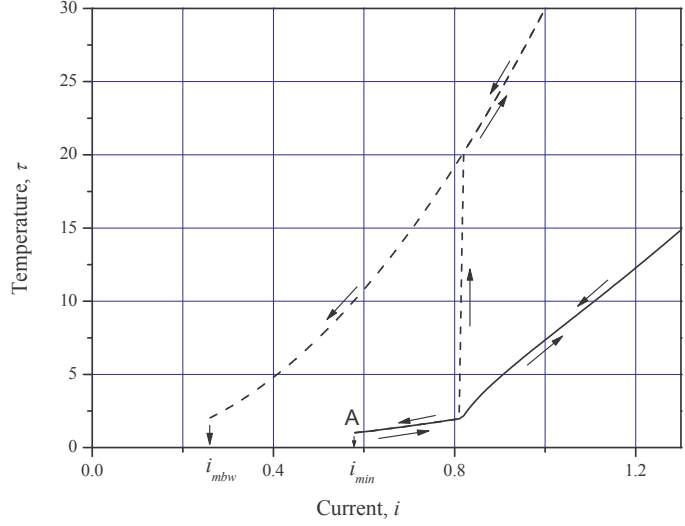

Fig. 6. Temperature in the normal state vs. current for a film (solid line) and a wire (dashed line). The boiling crisis temperature $\tau_{cr} = 2$, $h = 1$, $h_1 = 0.1$, $\alpha l = 3$, $\alpha_s = 3$.

*3.3. Wide films*

Let us consider the thermal stability of a wide film $l_s > l \gg 1$ assuming $l_s - l \gg 1$, $u_1 = u_0$ and $h = 1$. In this case the substrate temperature outside of the film ($y > l$) is given by a solution of Eq. (10):

$$\tau_1 = \tau_{10} e^{l-y}.$$

where $\tau_{10}$ is the temperature at the point $y = l$ and is determined from the boundary conditions. Here and below we assume that task is symmetrical and consider only a half of the structure at $y > 0$.

To solve Eq. (9), we use the condition that a voltage drop $u$ is the same in any point across the film. Using expression (6) a local current density $j$ can be presented as a function of $u$ and of a local film temperature $\tau_f$:

$$j = \begin{cases} (u/u_0)^{1/n}(1-\tau_f) & \text{for FC regime} \\ u - u_0 + 1 - \tau_f & \text{for FF regime} \\ u/[1+\gamma(\tau_f-1)] \approx u[1-\gamma(\tau_f-1)] & \text{for normal state} \end{cases} \quad (22)$$

Simplification used by us in the last expression for the normal state is valid at $\gamma\tau_f \ll 1$ and allows us to obtain an analytical solution. For all the regimes, the solution of Eq. (9) with the boundary conditions (11) and (13) has a form

$$\tau_f = \tau_0\left[1 - \frac{\cosh(\chi y)}{\cosh(\chi l) + \chi\sinh(\chi l)}\right], \quad (23)$$

where the values $\tau_0$ and $\chi$ are functions of $u$ presented in Tabl. 3 for different regimes.

Table 3

Explicit forms for $\tau_0$ and $\chi$ in solution (23)

| | $\tau_0$ | $\chi$ |
|---|---|---|
| FC regime | $\dfrac{\alpha(u/u_0)^{1/n} u}{1+\alpha(u/u_0)^{1/n} u}$ | $\sqrt{1+\alpha(u/u_0)^{1/n} u}$ |
| FF regime | $\dfrac{\alpha u (1+u-u_0)}{1+\alpha u}$ | $\sqrt{1+\alpha u}$ |
| Normal state | $\dfrac{\alpha u^2 (1+\gamma)}{1+\gamma\alpha u^2}$ | $\sqrt{1+\gamma\alpha u^2}$ |

For every regime the voltage drop $u$ is determined as a solution of the corresponding transcedental equation:
for the FC:

$$\left(\frac{u}{u_0}\right)^{1/n}\left[l(1-\tau_0)+\frac{\tau_0\sinh(\chi l)}{\chi\cosh(\chi l)+\chi^2\sinh(\chi l)}\right]=il, \tag{24}$$

for the FF:

$$\frac{u-u_0+1}{1+\alpha u}\left[l-\frac{\alpha u}{1+\alpha u+\chi\coth(\chi l)}\right]=il, \tag{25}$$

for the normal state

$$u\left\{[1+\gamma(1-\tau_0)]l+\frac{\tau_0\gamma\sinh(\chi l)}{\chi[\cosh(\chi l)+\chi\sinh(\chi l)]}\right\}=il. \tag{26}$$

For all the regimes the film temperature decreases and the current density increases monotonically with the coordinate $y$. As an example, the distributions of the temperature and current density across the film in the FC regime are presented in Fig. 7. Note also that for the FC regime we can obtain an analytical approximation for $\alpha u_0 \ll 1$ resulting in $\tau_0 \ll 1$. Then from (24) we have $u \approx u_0 i^n$ and expressions (22), (23) reduce to:

$$\tau_f \approx \alpha u_0 i^{n+1}\left[1-\frac{\cosh(y)}{\cosh(l)+\sinh(l)}\right],$$

$$j = i\left\{1-\alpha u_0 i^{n+1}\left[1-\frac{\cosh(y)}{\cosh(l)+\sinh(l)}\right]\right\}.$$

The last expressions clearly illustrate the dependences of the temperature and current density on the coordinate.

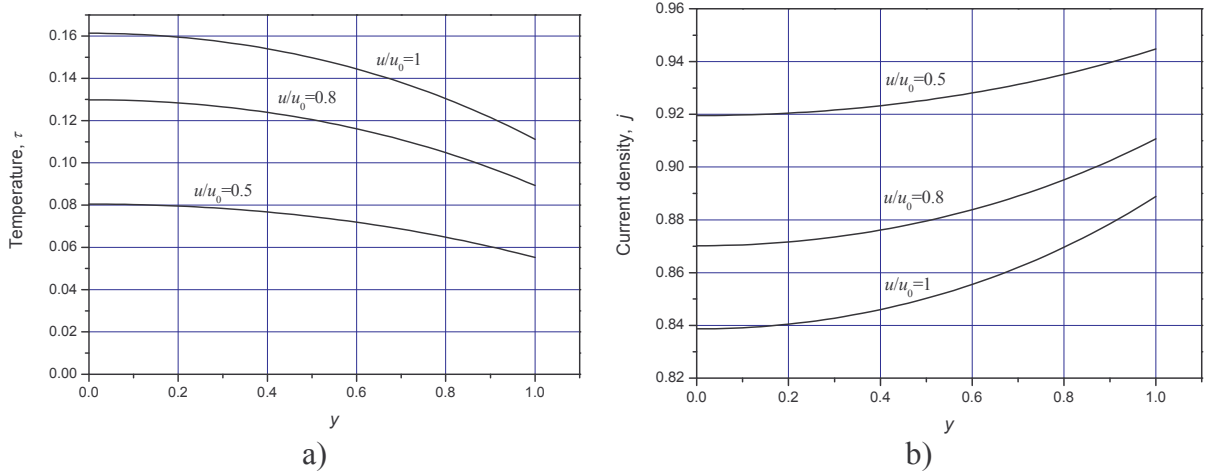

Fig. 7. Temperature (a) and current density (b) in the film vs. *y*-coordinate in the FC regime for $\alpha = 3$, $u_0 = 0.1$, $l=1$, $n=8$.

Similarly to a narrow film (Fig. 4), there are two types of a VCC in the FC regime: $u(i)$ is a single-valued function and $u(i)$ possesses two values at $i_1 < i < i_2$. The boundary value of $\alpha$, which separates these two types, depends on $n$, $u_0$ and $l$. For example, at $l = 1$, $n = 8$, $u_0 = 0.1$, the boundary value of $\alpha$ is about 2.8. The current $i_1$ can be determined from Eq. (24) substituting $u = u_0$:

$$i_1 = \frac{1}{1+\alpha u_0} + \frac{\alpha u_0}{l(1+\alpha u_0)\sqrt{1+\alpha u_0}} \frac{\sinh\left(l\sqrt{1+\alpha u_0}\right)}{\left[\cosh\left(l\sqrt{1+\alpha u_0}\right) + \sqrt{1+\alpha u_0}\,\sinh\left(l\sqrt{1+\alpha u_0}\right)\right]}.$$

The FF regime starts at a current $i_1$ and continues till $\tau_f = 1$, i. e. till the appearance of the normal state. However, in wide films, the situation is possible when at the same time the temperature in the film center is above the critical value, $\tau_f(y=0) > 1$ and the temperature at the film boundary is below the critical value, $\tau_f(y=l) < 1$. In this "mixed" regime there are both a zone of the normal state and a zone of the FF regime. The boundary between the zones is directed along a film, parallel to the *x*-axis.

Here we obtain a solution for the temperature distribution in the "mixed" regime for $\gamma = 0$ and $u_0 = 0$. In this case a current density $j = u$ in the normal state zone ($y < l_n$) and $j = u+1-\tau_f$ in the FF zone ($l_n < y < l$) (here $l_n$ is the width of the normal zone). The solutions of Eqs. (9) and (10) are

$$\tau_f = \begin{cases} \alpha u^2 + A_n \cosh(y) & \text{at } y \leq l_n \\ \dfrac{\alpha u(u+1)}{\alpha u+1} + A_f e^{-\chi y} + B_f e^{\chi y} & \text{at } l_n < y \leq l \end{cases}$$

$$\tau_1 = A_1 e^{-y} \quad \text{at } y > l$$

where $\chi = \sqrt{1+\alpha u}$; $u$, $l_n$ and the constants of integration are determined using Eq. (8), the continuity conditions of temperature and heat flux at $y = l_n$ and $y = l$, and the condition: $\tau_f(y = l_n) = 1$. The problem is reduced to solving the following system:

$$\tau_0 + (1-\tau_0)[\cosh(z) + \chi\sinh(z)] + \frac{1-\alpha u^2}{\chi}\tanh(l_n)[\chi\cosh(z) - \sinh(z)] = 0 \qquad (27)$$

$$ul + (1-\tau_0)(l - l_n) - \frac{1-\tau_0}{\chi}\sinh(z) + \frac{1-\alpha u^2}{\chi^2}\tanh(l_n)[\cosh(z) - 1] = il \qquad (28)$$

where $\tau_0 = \dfrac{\alpha u(1+u)}{1+\alpha u}$; $z = \chi(l - l_n)$.

The VCCs of wide films (Fig. 8) are similar to the VCCs presented in Fig. 2. However, the existence of the mixed regime leads to appearance of new portions between the solid straight and dashed lines. The solid straight line drawn through the coordinate origin gives the VCC of the film in the normal state, when the temperature of the film is above the critical value ($\tau_f \geq 1$) at any point. The dashed line corresponds to the state when the temperature in the film center achieves the critical value ($\tau_f(y=0) = 1$). Below this line there is only the FF regime. Note that even at $\alpha > 1$ thermal state of a superconductor in the mixed regime is stable.

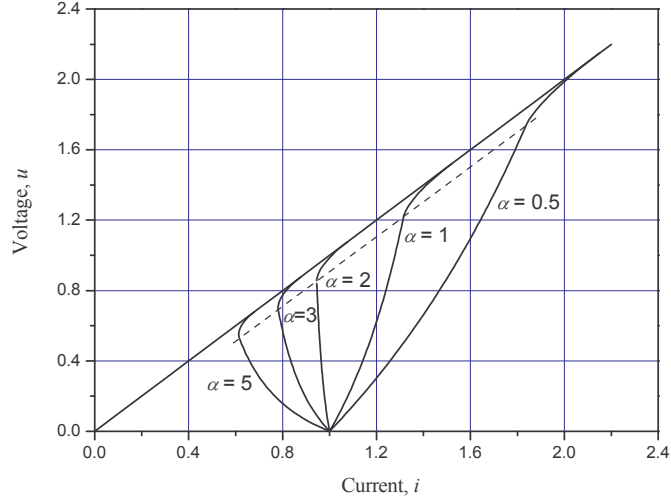

Fig. 8 Voltage-current characteristic of a wide film, $l = 1$. A straight line drawn through the coordinate origin gives the VCC of the film in the normal state. The zone between the solid and dashed straight lines is a region of the mixed regime.

In the general case a similar situation can be realized where the FF and FC regimes appear at the same time and also be separated by a line parallel to the x-axis. In principle, the regime at which there are simultaneously all three zones (FC, FF and normal) can exist. However, in the framework of the model where $u_1$ is independent of a temperature, $u_1 = u_0$, a regime, at which there are simultaneously the FF and FC zones, cannot be obtained.

## 4. Film with longitudinal thermal gradient

Above we have considered the stationary thermal state of a superconducting film with a uniform temperature distribution along the film. To investigate the thermal state of a film with nonuniform distribution of the temperature along a narrow film ($l \ll 1$) on a wide substrate ($l_s \gg 1$), we use a stepwise voltage-current characteristic:

$$u = \begin{cases} 0, & i < 1 - \tau_f, \\ i, & i \geq 1 - \tau_f. \end{cases} \qquad (29)$$

This approximation is frequently used to obtain analytical expressions for the analysis of the normal zone propagation and thermal domains [12, 13]. In the case of a narrow film one can assume that the thermal state of the film is determined by the thermal processes only in the substrate. The film temperature equals to the substrate temperature at $y = 0$:

$$\tau_f(t,x) = \tau_1(t,x,y)\big|_{y=0}$$

and the substrate temperature is given by a solution of Eq. (5). If $v$ is a steady-state velocity of the normal zone, Eq. (5) can be rewritten as ($h = 1$):

$$-v\frac{\partial \tau_1}{\partial X} = \frac{\partial^2 \tau_1}{\partial X^2} + \frac{\partial^2 \tau_1}{\partial y^2} - \tau_1, \qquad (30)$$

where $X = x-vt$ and the normal zone occupies the film at $X<0$. Due to symmetry of the task it is enough to consider a half the substrate at $y > 0$. The boundary conditions for Eq. (30) are:

$\tau_1 = 0$ at $y \to \infty$ or $X \to \infty$;

$\dfrac{\partial \tau_1}{\partial y} = -\alpha_{ef} i^2$ at $y = 0$ and $X \leq 0$;

$\dfrac{\partial \tau_1}{\partial y} = 0$ at $y = 0$ and $X > 0$.

The temperature at $X \to -\infty$ is given by a solution of Eq. (30) where the derivatives with respect to $X$ equal to zero, $\alpha_{ef} = \alpha l$.

The velocity $v$ is determined from the condition that the current equals to the critical value at the boundary between the normal and superconducting zones:

$i = 1-\tau_f$ at $X = 0$ and $y = 0$.

The task is reduced to the Helmholtz equation:

$$\frac{\partial^2 w}{\partial X^2} + \frac{\partial^2 w}{\partial y^2} = \left(1 + \frac{v^2}{4}\right)w, \qquad (31)$$

where $\tau_1 = w \exp(-vX/2)$.

Using the solution for Eq. (31) presented in [15] we obtain the solution of Eq. (30) in the following form:

$$\tau_1(X,y) = \frac{\alpha_{ef} i^2}{\pi} \int_{-\infty}^{0} e^{\frac{v(\xi-X)}{2}} K_0\left(\sqrt{\left(1+\frac{v^2}{4}\right)\left[(X-\xi)^2 + y^2\right]}\right) d\xi, \qquad (32)$$

where $K_0$ is the modified Bessel function of the zero order. Note that at $x \to -\infty$ $K_0(x) \sim e^{-x}/\sqrt{2x/\pi}$ and the integral in (32) has a finite magnitude at any $v$.

The velocity of the normal zone propagation is determined as a solution of the following equation:

$$1 - i = \frac{\alpha_{ef} i^2}{\pi} \int_{-\infty}^{0} e^{\frac{v\xi}{2}} K_0 \left( \sqrt{\left(1 + \frac{v^2}{4}\right) \xi^2} \right) d\xi. \tag{33}$$

One of the important parameters is the minimal current $i_m$ of the zone propagation [12, 13], which is corresponding to $v = 0$. In this case the integral in Eq. (33) gives $\pi/2$ and this current is

$$i_m = \frac{\sqrt{1 + 2\alpha_{ef}} - 1}{\alpha_{ef}}. \tag{34}$$

Fig. 9 gives the temperature distribution for this case.

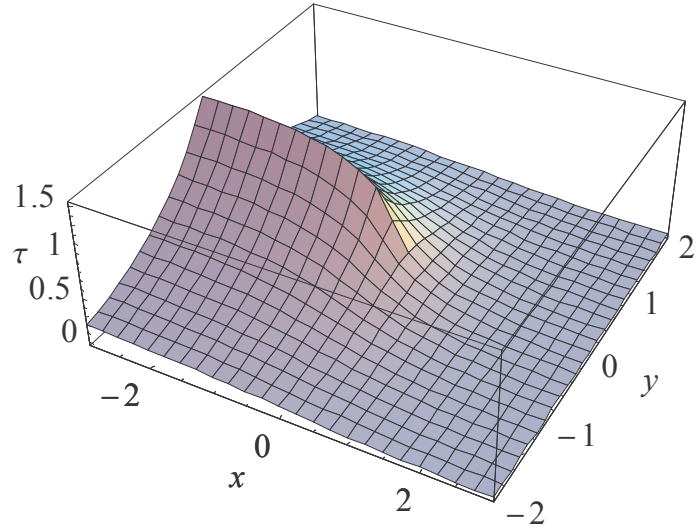

Fig. 9. Temperature distribution in the substrate at $\alpha_{ef} = 2$ and $i_m = 0.618$.

Expression (34) is congruent to the expression for the minimal current of the normal zone propagation along a superconducting wire [12], where the Stekly is replaced by an effective parameter $\alpha_{ef}$. However, the dependence of the velocity $v$ on a current in a film differs from that in wires (Fig. 10). In the case of a film the velocity is higher, especially, at a current above $i_m$.

## 5. Conclusion

The thermal state of a film deposited on a wide substrate was investigated in different regimes: flux creep, flux flow and normal. Using obtained analytical expressions describing VCCs and temperature-current dependence of a film, it was shown:

(a) For stationary problems such as obtaining the VCCs of narrow films without the boiling crisis, the determination of the minimum normal zone propagation current, one can introduce an effective parameter $\alpha_{ef}$, similar to the Stekly parameter. However, this parameter cannot be used for the cases involving the boiling crisis, wide films and the moving normal zones.

(b) Bistability of the thermal state of a film appears also in the FC regime. Moreover, there are the conditions when only the normal state exists in the stationary regime at a current below the critical one.

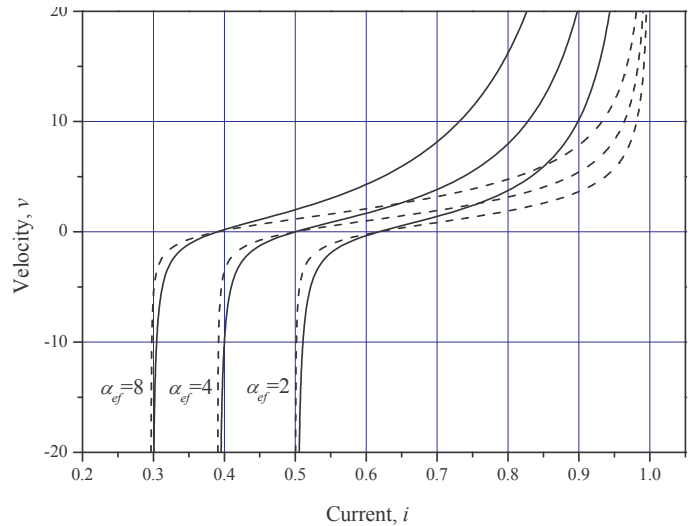

Fig. 10. Velocity of the normal zone propagation: solid line – velocity in a narrow film; dashed line – in a wire.

(c) Taking into account the FC, the boundary between stable and unstable states in the FF is shifted. The stable FF regime can be observed at $\alpha_{ef} > 1$ and a current below the critical value.

(d) Under the boiling crisis the thermal stability of films is higher than the stability of wires. The temperature-current dependence for a film differs from that for a wire and is characterized by a smooth temperature increase with the current. The film boiling zone on the substrate extends smoothly having the boundary with the nuclei boiling zone, while in the case of a wire the film boiling surrounds whole of the cooled surface.

(e) For wide films, the "mixed" regime can exist where the FF and the normal state are realized simultaneously with the boundary between them parallel to the film axis. This regime is thermally stable even at large values of the Stekly parameter.

Analysis of the propagation of a normal zone along a narrow film has shown:

(a) The minimum normal zone propagation current (the velocity $v = 0$) for films is determined as for wires using the effective Stekly parameter.

(b) The normal zone velocity in the film is higher than that in the wires. The last makes the films more attractive for application in superconducting switches and fault current limiters.